\documentclass[letterpaper, 10 pt, conference]{ieeeconf}  
\IEEEoverridecommandlockouts                              

\usepackage{graphics} 
\usepackage{epsfig} 
\usepackage{mathptmx} 
\usepackage{times} 
\usepackage{amsmath} 
\usepackage{amssymb}  
\usepackage{subfig}
\usepackage{algorithm,algorithmic}
\usepackage{amsfonts}
\usepackage{ntheorem}
\usepackage{hyperref}

\theoremstyle{break}
\newtheorem{theorem}{Theorem}
\newtheorem{assumption}{Assumption}
\usepackage{enumerate}
\usepackage{relsize}

\title{\LARGE \bf Model-Based Stochastic Search for Large Scale Optimization of Multi-Agent UAV Swarms}

\author{David D. Fan$^{1}$, Evangelos Theodorou$^{2}$, and John Reeder$^{3}$
\thanks{$^{1}$ Inst. for Robotics and Intelligent Machines, Georgia Institute of Technology. {\tt\small david.fan@gatech.edu}}%
\thanks{$^{2}$ Dept. of Aerospace Engineering, Georgia Institute of Technology. {\tt\small evangelos.theodorou@gatech.edu}}%
\thanks{$^{3}$ SPAWAR Systems Center Pacific, San Diego, CA, USA.
        {\tt\small john.d.reeder@navy.mil}}%
\thanks{$^{\dagger}$ Video at \url{http://goo.gl/dWvQi7}} %
}

\begin{document}

\maketitle
\thispagestyle{empty}
\pagestyle{empty}

\begin{abstract}
Recent work from the reinforcement learning community has shown that Evolution Strategies are a fast and scalable alternative to other reinforcement learning methods.  In this paper we show that Evolution Strategies are a special case of model-based stochastic search methods.  This class of algorithms has nice asymptotic convergence properties and known convergence rates.  We show how these methods can be used to solve both cooperative and competitive multi-agent problems in an efficient manner.  We demonstrate the effectiveness of this approach on two complex multi-agent UAV swarm combat scenarios: where a team of fixed wing aircraft must attack a well-defended base, and where two teams of agents go head to head to defeat each other$^{\dagger}$.
\end{abstract}

\section{Introduction}
Reinforcement Learning is concerned with maximizing rewards from an environment through repeated interactions and trial and error.  Such methods often rely on various approximations of the Bellman equation and include value function approximation, policy gradient methods, and more \cite{Li2017}.  The Evolutionary Computation community, on the other hand, have developed a suite of methods for black box optimization and heuristic search \cite{Stanley2005}.  Such methods have been used to optimize the structure of neural networks for vision tasks, for instance \cite{RN43}.
  
\begin{figure}[h]
\centering
\includegraphics[width=0.4\textwidth]{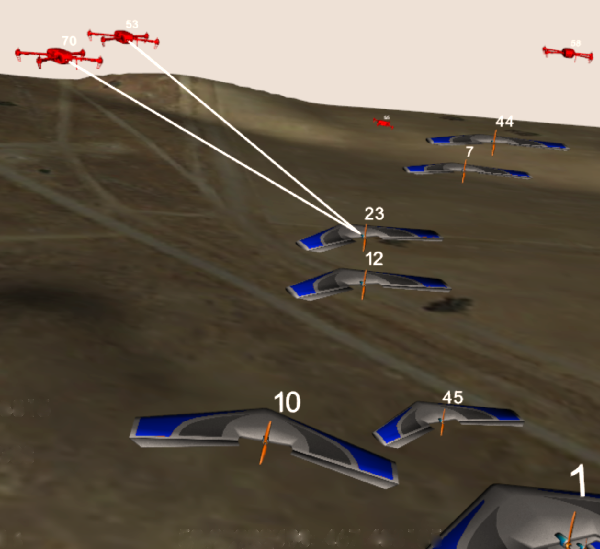}
\caption{The SCRIMMAGE multi-agent simulation environment.  In this scenario, blue team fixed-wing agents attack red team quadcopter defenders.  White lines indicate missed shots.}
\label{fig:splash}
\end{figure}

Recently, Salimans et al. have shown that a particular variant of evolutionary computation methods, termed Evolution Strategies (ES) are a fast and scalable alternative to other reinforcement learning approaches, solving the difficult humanoid MuJoCo task in 10 minutes \cite{Salimans2017}.  The authors argue that ES has several benefits over other reinforcement learning methods: 1) The need to backpropagate gradients through a policy is avoided, which opens up a wider class of policy parameterizations;  2) ES methods are massively parallelizable, which allows for scaling up learning to larger, more complex problems;  3)  ES often finds policies which are more robust than other reinforcement learning methods; and 4)  ES are better at assigning credit to changes in the policy over longer timescales, which enables solving tasks with longer time horizons and sparse rewards.  In this work we leverage all four of these advantages by using ES to solve a problem with: 1) a more complex and decipherable policy architecture which allows for safety considerations; 2) a large-scale simulated environment with many interacting elements; 3) multiple sources of stochasticity including variations in intial conditions, disturbances, etc.; and 4) sparse rewards which only occur at the very end of a long episode.

A common critique of evolutionary computation algorithms is a lack of convergence analysis or guarantees.  Of course, for problems with non-differentiable and non-convex objective functions, analysis will always be difficult.  Nevertheless, we show that the Evolution Strategies algorithm proposed by \cite{Salimans2017} is a special case of a class of model-based stochastic search methods known as Gradient-Based Adaptive Stochastic Search (GASS) \cite{hu2015model}.  This class of methods generalizes many stochastic search methods such as the well-known Cross Entropy Method (CEM) \cite{RN9}, CMA-ES \cite{DBLP:journals/corr/Hansen16a}, etc.  By casting a non-differentiable, non-convex optimization problem as a gradient descent problem, one can arrive at nice asymptotic convergence properties and known convergence rates \cite{zhou2014gradient}.  

With more confidence in the convergence of Evolution Strategies, we demonstrate how ES can be used to efficiently solve both cooperative and competitive large-scale multi-agent problems.  Many approaches to solving multi-agent problems rely on hand-designed and hand-tuned algorithms (see \cite{RN25} for a review).  One such example, distributed Model Predictive Control, relies on independent MPC controllers on each agent with some level of coordination between them \cite{RN35,RN8}. These controllers require hand-designing dynamics models, cost functions, feedback gains, etc. and require expert domain knowledge. Additionally, scaling these methods up to more complex problems continues to be an issue. Evolutionary algorithms have also been tried as a solution to multi-agent problems; usually with smaller, simpler environments, and policies with low complexity \cite{RN13,RN4}.  Recently, a hybrid approach combining MPC and the use of genetic algorithms to evolve the cost function for a hand-tuned MPC controller has been demonstrated for a UAV swarm combat scenario \cite{Fan2017}.

In this work we demonstrate the effectiveness of our approach on two complex multi-agent UAV swarm combat scenarios: where a team of fixed wing aircraft must attack a well-defended base, and where two teams of agents go head to head to defeat each other.  Such scenarios have been previously considered in simulated environments with less fidelity and complexity \cite{Gaerther2013,Fan2017}.  We leverage the computational efficiency and flexibility of the recently developed SCRIMMAGE multi-agent simulator for our experiments (Figure \ref{fig:splash}) \cite{demarco_scrimmage_2018}.  We compare the performance of ES against the Cross Entropy Method.  We also show for the competitive scenario how the policy learns over time to coordinate a strategy in response to an enemy learning to do the same.  We make our code freely available for use (\url{https://github.com/ddfan/swarm_evolve}).

\section{Problem Formulation}
We can pose our problem as the non-differentiable, non-convex optimization problem
\begin{equation}
\theta^*=\arg\max_{\theta\in\Theta}J(\theta)
\label{eq:1}
\end{equation}
where $\Theta\subset\mathbb{R}^n$, a nonempty compact set, is the space of solutions, and $J(\theta)$ is a non-differentiable, non-convex real-valued objective function $J:\Theta\rightarrow\mathbb{R}$.  $\theta$ could be any combination of decision variables of our problem, including neural network weights, PID gains, hardware design parameters, etc. which affect the outcome of the returns $J$.  For reinforcement learning problems $\theta$ usually represents the parameters of the policy and $J$ is an implicit function of the sequential application of the policy to the environment.  We first review how this problem can be solved using Gradient-Based Adaptive Stochastic Search methods and then show how the ES algorithm is a special case of these methods.

\subsection{Gradient-Based Adaptive Stochastic Search}
The goal of model-based stochastic search methods is to cast the non-differentiable optimization problem (\ref{eq:1}) as a differentiable one by specifying a probabilistic model (hence "model-based") from which to sample \cite{zhou2014gradient}.  Let this model be $p(\theta|\omega)=f(\theta;\omega), \omega\in\Omega$, where $\omega$ is a parameter which defines the probability distribution (e.g. for  Gaussian distributions, the distribution is fully parameterized by the mean and variance $\omega=[\mu,\sigma]$).  Then the expectation of $J(\theta)$ over the distribution $f(\theta;\omega)$ will always be less than the optimal value of $J$, i.e.
\begin{equation}
\int_{\Theta} J(\theta)f(\theta;\omega)d\theta\leq J(\theta^*)
\label{eq:2}
\end{equation}

The idea of Gradient-based Adaptive Stochastic Search (GASS) is that one can perform a search in the space of parameters of the distribution $\Omega$ rather than $\Theta$, for a distribution which maximizes the expectation in (\ref{eq:2}):
\begin{equation}
\label{eq:3}
\omega^*=\arg\max_{\omega\in\Omega}\int_{\Theta}J(\theta)f(\theta;\omega)d\theta
\end{equation}
Maximizing this expectation corresponds to finding a distribution which is maximally distributed around the optimal $\theta$.  However, unlike maximizing (\ref{eq:1}), this objective function can now be made continuous and differentiable with respect to $\omega$.  With some assumptions on the form of the distribution, the gradient with respect to $\omega$ can be pushed inside the expectation.  

The GASS algorithm presented by \cite{zhou2014gradient} is applicable to the exponential family of probability densities:
\begin{equation}
f(\theta;\omega)=\exp\{\omega^\intercal T(\theta)-\phi(\theta)\}
\end{equation}
where $\phi(\theta)=\ln\int\exp(\omega^\intercal T(\theta))d\theta$, and $T(\theta)$ is the vector of sufficient statistics.  Since we are concerned with showing the connection with ES which uses parameter perturbations sampled with Gaussian noise, we assume that $f(\theta;\omega)$ is Gaussian.  Furthermore, since we are concerned with learning a large number of parameters (i.e. weights in a neural network), we assume an independent Gaussian distribution over each parameter.  Then, $T(\theta)=[\theta,\theta^2]^\intercal\in\mathbb{R}^{2n}$ and $\omega=[\mu/\sigma^2,-1/n\sigma^2]^\intercal\in\mathbb{R}^{2n}$, where $\mu$ and $\sigma$ are vectors of the mean and standard deviation corresponding to the distribution of each parameter, respectively. 

\begin{algorithm}
\caption{Gradient-Based Adaptive Stochastic Search}
\label{alg:gass}
\begin{algorithmic}[1]
  \REQUIRE Learning rate $\alpha_k$, sample size $N_k$, initial policy parameters $\omega_0=[\mu_0,\sigma_0^2]^\intercal$, smoothing function $S()$, small constant $\gamma>0$.
  \FOR{$k=0,1,\cdots$}
  	\STATE Sample $\theta_k^i\stackrel{iid}{\sim}f(\theta;\omega_k),i=1,2,\cdots,N_k$.
  	\STATE Compute returns $w_k^i=S(J(\theta_k^i))$ for $i=1,\cdots,N_k$.
  	\STATE Compute variance terms $V_k=\hat{V}_k+\gamma I$, eq (\ref{eq:var}),(\ref{eq:var2}) 
  	\STATE Calculate normalizer $\eta=\sum_{i=1}^{N_k}w_k^i$.
  	\STATE Update $\omega_{k+1}$:
  	\STATE $\omega_{k+1}\leftarrow\omega_k+\alpha_k\frac{1}{\eta}V_k^{-1}\mathlarger{\sum}\limits_{i=1}^{N_k}w_k^i\bigg(\begin{bmatrix} \theta_k^i\\(\theta_k^i)^2 \end{bmatrix}-\begin{bmatrix} \mu\\\sigma^2+\mu^2 \end{bmatrix}\bigg)$
  \ENDFOR
\end{algorithmic}
\end{algorithm}

We present the GASS algorithm for this specific set of probability models (Algorithm \ref{alg:gass}), although the analysis for convergence holds for the more general exponential family of distributions.  For each iteration $k$, The GASS algorithm involves drawing $N_k$ samples of parameters $\theta_k^i\stackrel{iid}{\sim}f(\theta;\omega_k),i=1,2,\cdots,N_k$.  These parameters are then used to sample the return function $J(\theta_k^i)$.  The returns are fed through a shaping function $S(\cdot):\mathbb{R}\rightarrow\mathbb{R}^+$ and then used to calculate an update on the model parameters $\omega_{k+1}$. 

The shaping function $S(\cdot)$ is required to be nondecreasing and bounded from above and below for bounded inputs, with the lower bound away from 0.  Additionally, the set $\{\arg\max_{\theta\in\Theta}S(J(\theta))\}$ must be a nonempty subset of the set of solutions of the original problem $\{\arg\max_{\theta\in\Theta}J(\theta)\}$.  The shaping function can be used to adjust the exploration/exploitation trade-off or help deal with outliers when sampling.  The original analysis of GASS assumes a more general form of $S_k(\cdot)$ where $S$ can change at each iteration.  For simplicity we assume here it is deterministic and unchanging per iteration.

GASS can be considered a second-order gradient method and requires estimating the variance of the sampled parameters:
\begin{multline}
\hat{V}_k=\frac{1}{N_k-1}\sum_{i=1}^{N_k}T(\theta_k^i)T(\theta_k^i)^\intercal\\
-\frac{1}{N_k^2-N_k}\Bigg(\sum_{i=1}^{N_k}T(\theta_k^i)\Bigg)\Bigg(\sum_{i=1}^{N_k}T(\theta_k^i)\Bigg)^\intercal.
\label{eq:var}
\end{multline}
In practice if the size of the parameter space $\Theta$ is large, as is the case in neural networks, this variance matrix will be of size $2n\times 2n$ and will be costly to compute.  In our work we approximate $\hat{V}_k$ with independent calculations of the variance on the parameters of each independent Gaussian.  With a slight abuse of notation, consider $\tilde{\theta}_k^i$ as a scalar element of $\theta_k^i$.  We then have, for each scalar element $\tilde{\theta}_k^i$ a $2\times 2$ variance matrix:

\begin{multline}
\hat{V}_k=\frac{1}{N_k-1}\sum_{i=1}^{N_k}\begin{bmatrix} \tilde{\theta}_k^i\\(\tilde{\theta}_k^i)^2\end{bmatrix}\begin{bmatrix} \tilde{\theta}_k^i&(\tilde{\theta}_k^i)^2\end{bmatrix}\\
-\frac{1}{N_k^2-N_k}\Bigg(\sum_{i=1}^{N_k}\begin{bmatrix} \tilde{\theta}_k^i\\(\tilde{\theta}_k^i)^2\end{bmatrix}\Bigg)\Bigg(\sum_{i=1}^{N_k}\begin{bmatrix} \tilde{\theta}_k^i&(\tilde{\theta}_k^i)^2\end{bmatrix}\Bigg).
\label{eq:var2}
\end{multline}

Theorem \ref{thm:1} shows that GASS produces a sequence of $\omega_k$ that converges to a limit set which specifies a set of distributions that maximize (\ref{eq:3}).  Distributions in this set will specify how to choose $\theta^*$ to ultimately maximize (\ref{eq:1}).  As with most non-convex optimization algorithms, we are not guaranteed to arrive at the global maximum, but using probabilistic models and careful choice of the shaping function should help avoid early convergence into suboptimal local maximum.  The proof relies on casting the update rule in the form of a generalized Robbins-Monro algorithm (see \cite{zhou2014gradient}, Thms 1 and 2).  Theorem \ref{thm:1} also specifies convergence rates in terms of the number of iterations $k$, the number of samples per iteration $N_k$, and the learning rate $\alpha_k$.  In practice Theorem \ref{thm:1} implies the need to carefully balance the increase in the number of samples per iteration and the decrease in learning rate as iterations progress.

\begin{assumption}
\begin{enumerate}[i)]
\item The learning rate $\alpha_k>0$, $\alpha_k\rightarrow 0$ as $k\rightarrow\infty$, and $\sum_{k=0}^\infty \alpha_k=\infty$.
\item The sample size $N_k=N_0k^\xi$, where $\xi>0$; also $\alpha_k$ and $N_k$ jointly satisfy $\alpha/\sqrt{N_k}=\mathcal{O}(k^{-\beta})$.
\item $T(\theta)$ is bounded on $\Theta$.
\item If $\omega^*$ is a local maximum of (\ref{eq:3}), the Hessian of $\int_{\Theta}J(\theta)f(\theta;\omega)d\theta$ is continuous and symmetric negative definite in a neighborhood of $\omega^*$.
\end{enumerate}
\label{ass:1}
\end{assumption}
\begin{theorem}
Assume that Assumption \ref{ass:1} holds.  Let $\alpha_k=\alpha_0/k^\alpha$ for $0<\alpha<1$.  Let $N_k=N_0k^{\tau-\alpha}$ where $\tau>2\alpha$ is a constant.  Then the sequence $\{\omega_k\}$ generated by Algorithm \ref{alg:gass} converges to a limit set w.p.1. with rate $\mathcal{O}(1/\sqrt{k^\tau})$.
\label{thm:1}
\end{theorem}
\subsection{Evolutionary Strategies}
We now review the ES algorithm proposed by \cite{Salimans2017} and show how it is a first-order approximation of the GASS algorithm.  The ES algorithm consists of the same two phases as GASS:  1)  Randomly perturb parameters with noise sampled from a Gaussian distribution.  2) Calculate returns and calculate an update to the parameters.  The algorithm is outlined in Algorithm \ref{alg:es}.  Once returns are calculated, they are sent through a function $S(\cdot)$ which performs fitness shaping \cite{wierstra2014}.  Salimans et al. used a rank transformation function for $S(\cdot)$ which they argue reduced the influence of outliers at each iteration and helped to avoid local optima.
\begin{algorithm}
\caption{Evolution Strategies}
\label{alg:es}
\begin{algorithmic}[1]
  \REQUIRE Learning rate $\alpha_k$, noise standard deviation $\gamma$, initial policy parameters $\theta_0$, smoothing function $S()$.
  \FOR{$k=0,1,\cdots$}
  	\STATE Sample $\epsilon_1,\cdots,\epsilon_n\sim\mathcal{N}(0,I^{n\times n})$
  	\STATE Compute returns $w_k^i=S(J(\theta_k+\gamma\epsilon_i))$ for $i=1,\cdots,N_k$
  	\STATE Update $\theta_{k+1}$:
  	\STATE $\theta_{k+1}\leftarrow\theta_k+\alpha_k\frac{1}{N_k\gamma}\sum_{i=1}^{N_k}w_k^i\epsilon_i$
  \ENDFOR
\end{algorithmic}
\end{algorithm}
It is clear that the ES algorithm is a sub-case of the GASS algorithm when the sampling distribution is a point distribution.  We can also recover the ES algorithm by ignoring the variance terms on line $7$ in Algorithm \ref{alg:gass}.  Instead of the normalizing term $\eta$, ES uses the number of samples $N_k$.  The small constant in GASS $\gamma$ becomes the variance term in the ES algorithm.  The update rule in Algorithm \ref{alg:es} involves multiplying the scaled returns by the noise, which is exactly $\theta_k^i-\mu$ in Algorithm \ref{alg:gass}.
\begin{figure*}[ht]
\centering
\includegraphics[trim={0 9.5cm 4cm 0},clip,width=1.0\textwidth]{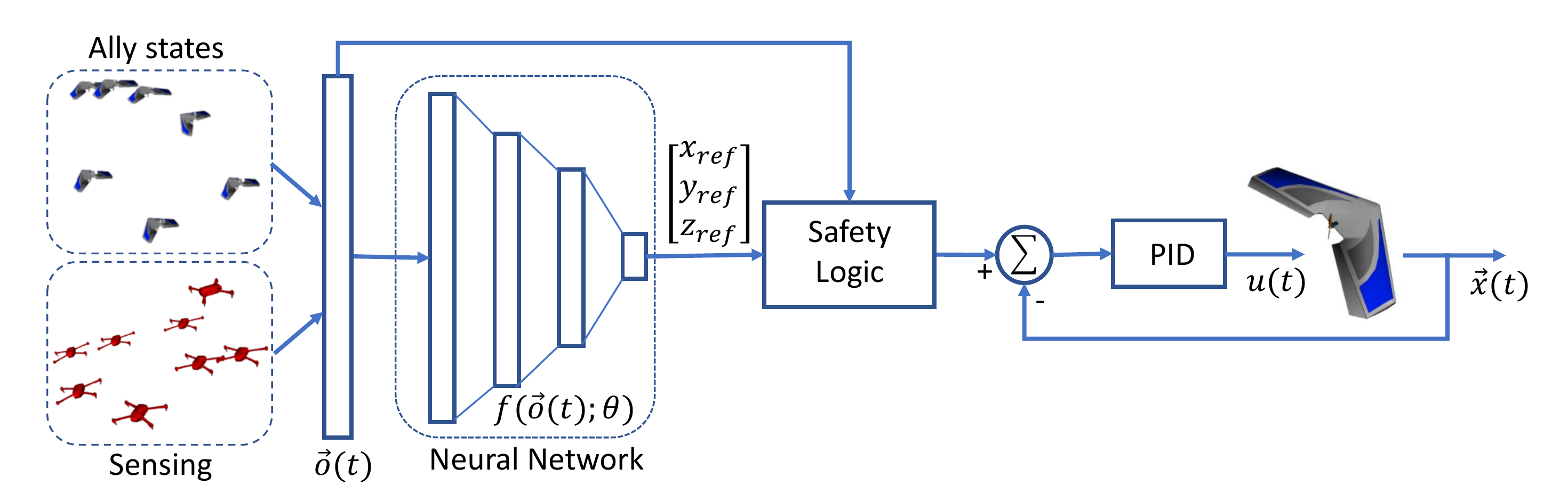}
\caption{Diagram of each agent's policy.  Nearby ally states and sensed enemies, base locations, etc. along with the agent's own state are fed into a neural network which produces a reference target in relative xyz coordinates.  The target is fed into the safety logic block which checks for collisions with neighbors or the ground.  It produces a reference target which is fed to the PID controller, which in turn provides low-level controls for the agent (thrust, aileron, elevator, rudder).}
\label{fig:policy}
\end{figure*}
We see that ES enjoys the same asymptotic convergence rates offered by the analysis of GASS.  While GASS is a second-order method and ES is only a first-order method, in practice ES uses approximate second-order gradient descent methods which adapt the learning rate in order to speed up and stabilize learning.  Examples of these methods include ADAM, RMSProp, SGD with momentum, etc., which have been shown to perform very well for neural networks.  Therefore we can treat ES a first-order approximation of the full second-order variance updates which GASS uses.  In our experiments we use ADAM \cite{kingma2014adam} to adapt the learning rate for each parameter.  As similarly reported in \cite{Salimans2017}, when using adaptive learning rates we found little improvement over adapting the variance of the sampling distribution.  We hypothesize that a first order method with adaptive learning rates is sufficient for achieving good performance when optimizing neural networks.  For other types of policy parameterizations however, the full second-order treatment of GASS may be more useful.  It is also possible to mix and match which parameters require a full variance update and which can be updated with a first-order approximate method.  We use the rank transformation function for $S(\cdot)$ and keep $N_k$ constant.

\subsection{Learning Structured Policies for Multi-Agent Problems}
Now that we are more confident about the convergence of the ES/GASS method, we show how ES can be used to optimize a complex policy in a large-scale multi-agent environment.  We use the SCRIMMAGE multi-agent simulation environment \cite{demarco_scrimmage_2018} as it allows us to quickly and in parallel simulate complex multi-agent scenarios.  We populate our simulation with 6DoF fixed-wing aircraft and quadcopters with dynamics models having 10 and 12 states, respectively.  These dynamcis models allow for full ranges of motion within realistic operating regimes.  Stochastic disturbances in the form of wind and control noise are modeled as additive Gaussian noise.  Ground and mid-air collisions can occur which result in the aircraft being destroyed.  We also incorporate a weapons module which allows for targeting and firing at an enemy within a fixed cone projecting from the aircraft's nose.  The probability of a hit depends on the distance to the target and the total area presented by the target to the attacker.  This area is based on the wireframe model of the aircraft and its relative pose.  For more details, see our code and the SCRIMMAGE simulator documentation.

We consider the case where each agent uses its own policy to compute its own controls, but where the parameters of the policies are the same for all agents. This allows each agent to control itself in a decentralized manner, while allowing for beneficial group behaviors to emerge.  Furthermore, we assume that friendly agents can communicate to share states with each other (see Figure \ref{fig:policy}).  Because we have a large number of agents (up to 50 per team), to keep communication costs lower we only allow agents to share information locally, i.e. agents close to each other have access to each other's states.  In our experiments we allow each agent to sense the states of the closest 5 friendly agents for a total of $5*10=50$ incoming state messages.

Additionally, each agent is equipped with sensors to detect enemy agents.  Full state observability is not available here, instead we assume that sensors are capable of sensing an enemy's relative position and velocity.  In our experiments we assumed that each agent is able to sense the nearest 5 enemies for a total of $5*7=35$ dimensions of enemy data ($7$ states = [relative xyz position, distance, and relative xyz velocities]).  The sensors also provide information about home and enemy base relative headings and distances (an additional $8$ states).  With the addition of the agent's own state ($9$ states), the policy's observation input $\vec{o}(t)$ has a dimension of $102$.  These input states are fed into the agent's policy: a neural network $f(\vec{o}(t);\theta)$ with 3 fully connected layers with sizes 200, 200, and 50, which outputs 3 numbers representing a desired relative heading $[x_{ref},y_{ref},z_{ref}]$.  Each agent's neural network has more than 70,000 parameters.  Each agent uses the same neural network parameters as its teammates, but since each agent encounters a different observation at each timestep, the output of each agent's neural network policy will be unique.  It may also be possible to learn unique policies for each agent; we leave this for future work.

With safety being a large concern in UAV flight, we design the policy to take into account safety and control considerations.  The relative heading output from the neural network policy is intended to be used by a PID controller to track the heading.  The PID controller provides low-level control commands $u(t)$ to the aircraft (thrust, aileron, elevator, rudder).  However, to prevent cases where the neural network policy guides the aircraft into crashing into the ground or allies, etc., we override the neural network heading with an avoidance heading if the aircraft is about to collide with something.  This helps to focus the learning process on how to intelligently interact with the environment and allies rather than learning how to avoid obvious mistakes.  Furthermore, by designing the policy in a structured and interpretable way, it will be easier to take the learned policy directly from simulation into the real world.  Since the neural network component of the policy does not produce low-level commands, it is invariant to different low-level controllers, dynamics, PID gains, etc.  This aids in learning more transferrable policies for real-world applications.
\section{Experiments}
We consider two scenarios:  a base attack scenario where a team of 50 fixed wing aircraft must attack an enemy base defended by 20 quadcopters, and a team competitive task where two teams concurrently learn to defeat each other.  In both tasks we use the following reward function:
\begin{multline}
J=10\times (\text{\#kills}) + 50\times(\text{\#collisions with enemy base})\\
 - 1e-5\times(\text{distance from enemy base at end of episode})
\end{multline}
The reward function encourages air-to-air combat, as well as suicide attacks against the enemy base (e.g. a swarm of cheap, disposable drones carrying payloads).  The last term encourages the aircraft to move towards the enemy during the initial phases of learning.

\subsection{Base Attack Task}
\begin{figure}[h]
\centering
\includegraphics[trim={0 2cm 0 1cm},clip,width=0.45\textwidth]{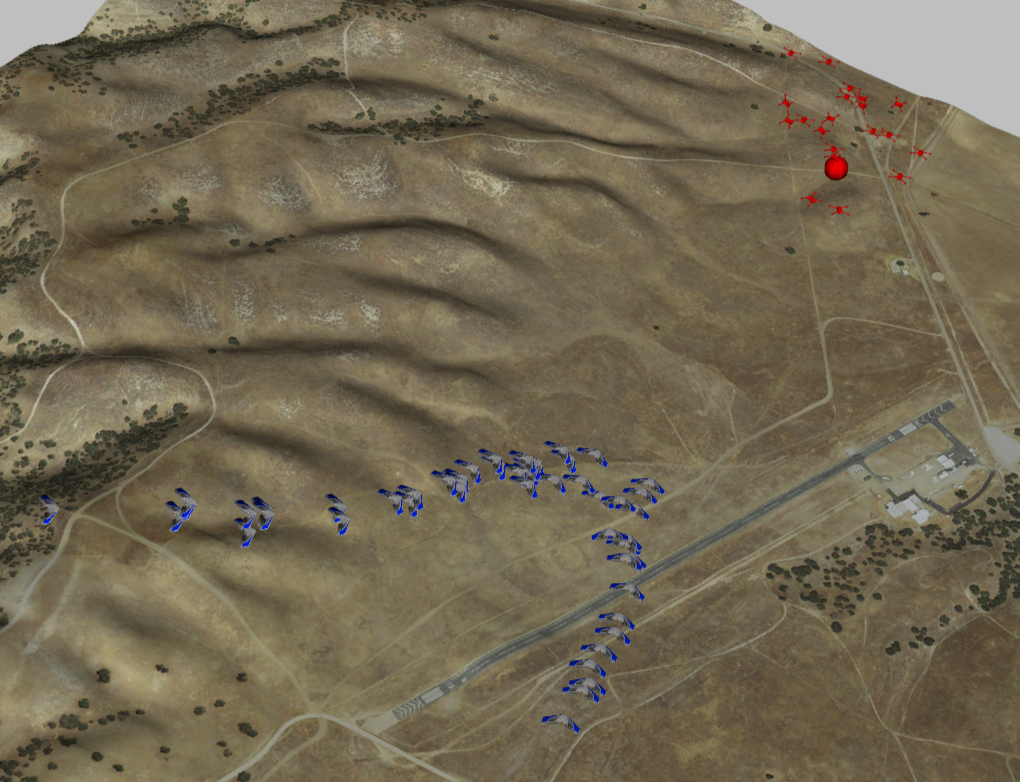}
\caption{Snapshot of base attack task.  The goal of the blue fixed wing team (lower left) is to attack the red base (red dot, upper right) while avoiding or attacking red quadcopter guards.}
\label{fig:baseattack}
\end{figure}

\begin{figure}[h]
  \centering
  \subfloat[Training]{\includegraphics[width=1.0\linewidth]{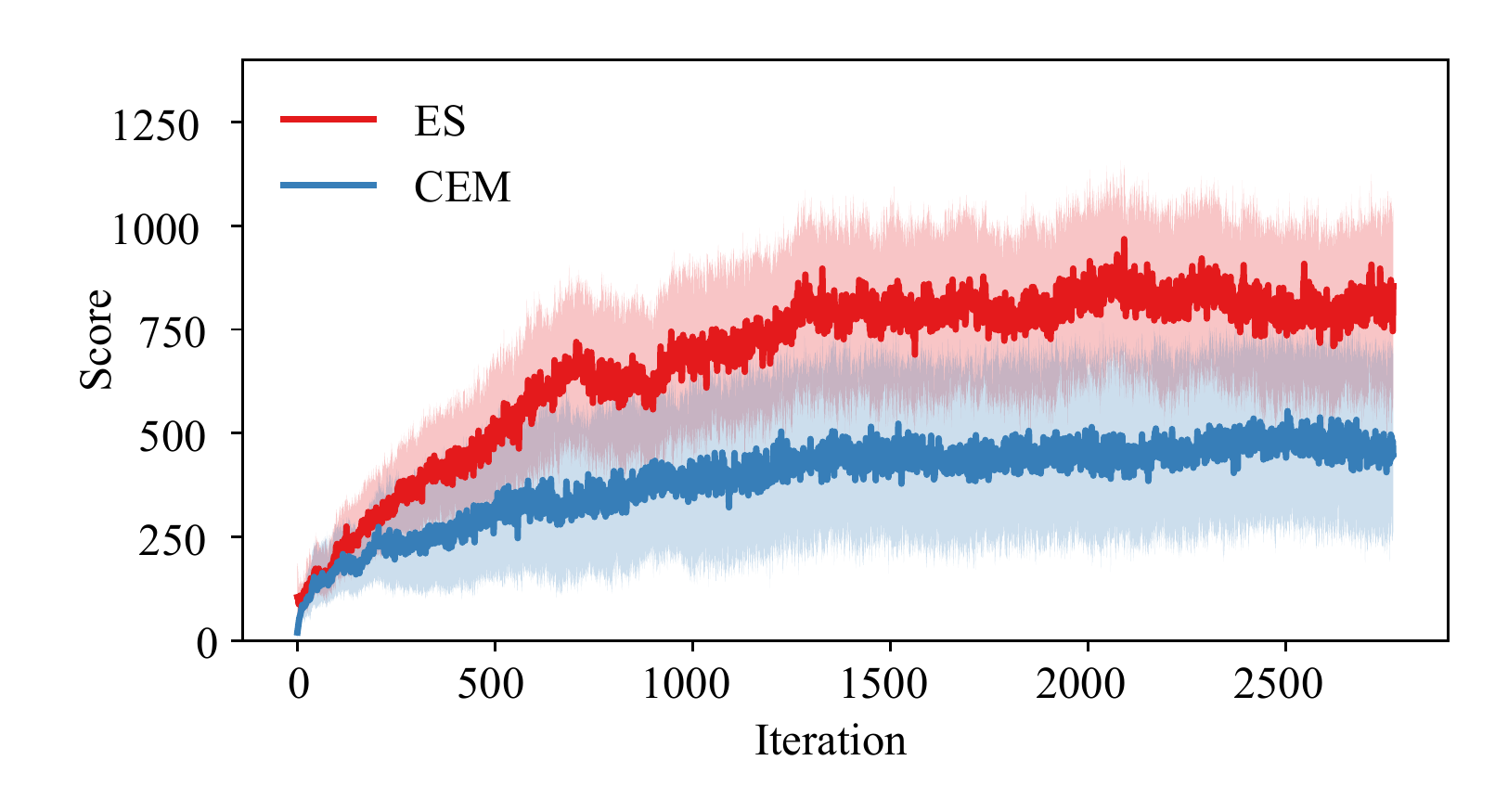}}\\\vspace*{-1em}
  \subfloat[Testing]{\includegraphics[width=1.0\linewidth]{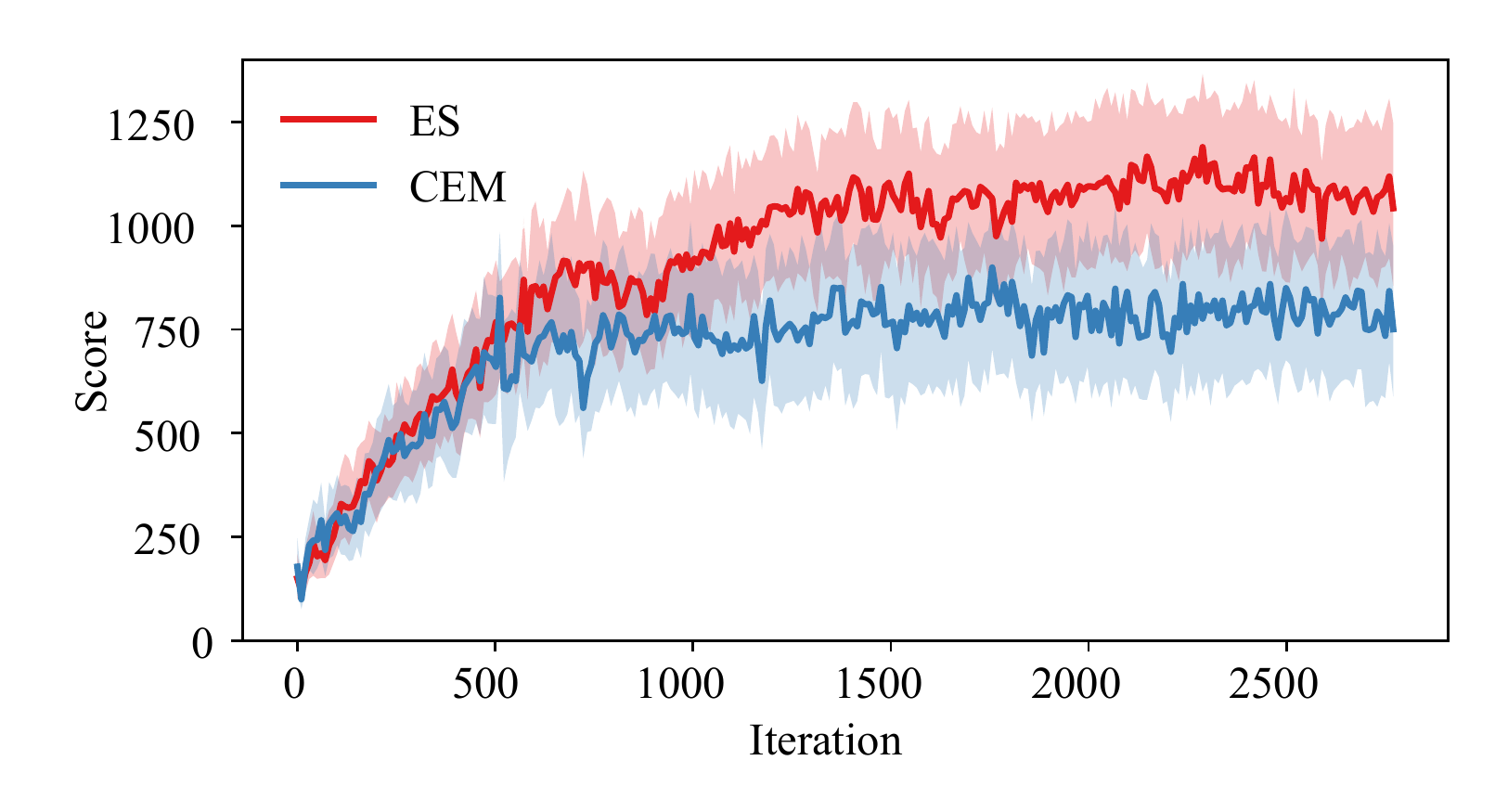}}
  \caption{Scores per iteration for the base attack task.  Top: Scores earned by perturbed policies during training.  Scores are on average lower because they result from policies which are parameterized by randomly peturbed values.  Bottom:  Scores during the course of training earned by the updated policy parameters.  Red curve is Evolution Strategies algorithm, blue is Cross Entropy Method.  Bold line is the median, shaded areas are 25/75 quartile bounds.}
   \label{fig:baseattack_learn}
\end{figure}
In this scenario a team of 50 fixed-wing aircraft must attack an enemy base defended by 20 quadcopters (Figure \ref{fig:baseattack}).  The quadcopters use a hand-crafted policy where in the absence of an enemy, they spread themselves out evenly to cover the base.  In the presence of an enemy they target the closest enemy, match that enemy's altitude, and fire repeatedly.  We used $N_k=300, \gamma=0.02$, a time step of $0.1$ seconds, and total episode length of $200$ seconds.   Initial positions of both teams were randomized in a fixed area at opposide ends of the arena.  Training took two days with full parallelization on a machine equipped with a Xeon Phi CPU (244 threads).

We found that over the course of training the fixed-wing team learned a policy where they quickly form a V-formation and approach the base.  Some aircraft suicide-attack the enemy base while others begin dog-fighting (see supplementary video\footnote{\url{http://https://goo.gl/dWvQi7}}).  We also compared our implementation of the ES method against the well-known cross-entropy method (CEM).  CEM performs significantly worse than ES (Figure \ref{fig:baseattack_learn}).  We hypothesize this is because CEM throws out a significant fraction of sampled parameters and therefore obtains a worse estimate of the gradient of (\ref{eq:3}).  Comparison against other full second-order methods such as CMA-ES or the full second-order GASS algorithm is unrealistic due to the large number of parameters in the neural network and the prohibitive computational difficulties with computing the covariances of those parameters.
\subsection{Two Team Competitive Match}
\begin{figure}[h]
\centering
\includegraphics[width=0.45\textwidth]{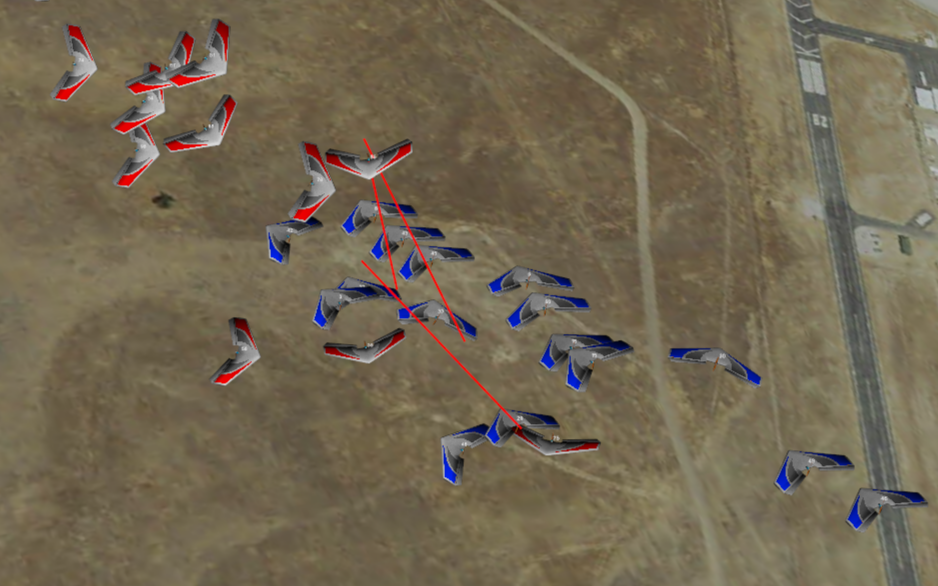}
\caption{Snapshot of two team competitive match.  The goal of both teams is to defeat all enemy planes while suffering minimum losses, or to attack the opponent's base.  Red lines indicate successful firing hit.}
\label{fig:versus}
\end{figure}
\begin{figure}[h]
  \centering
  \subfloat[Training]{\includegraphics[width=1.0\linewidth]{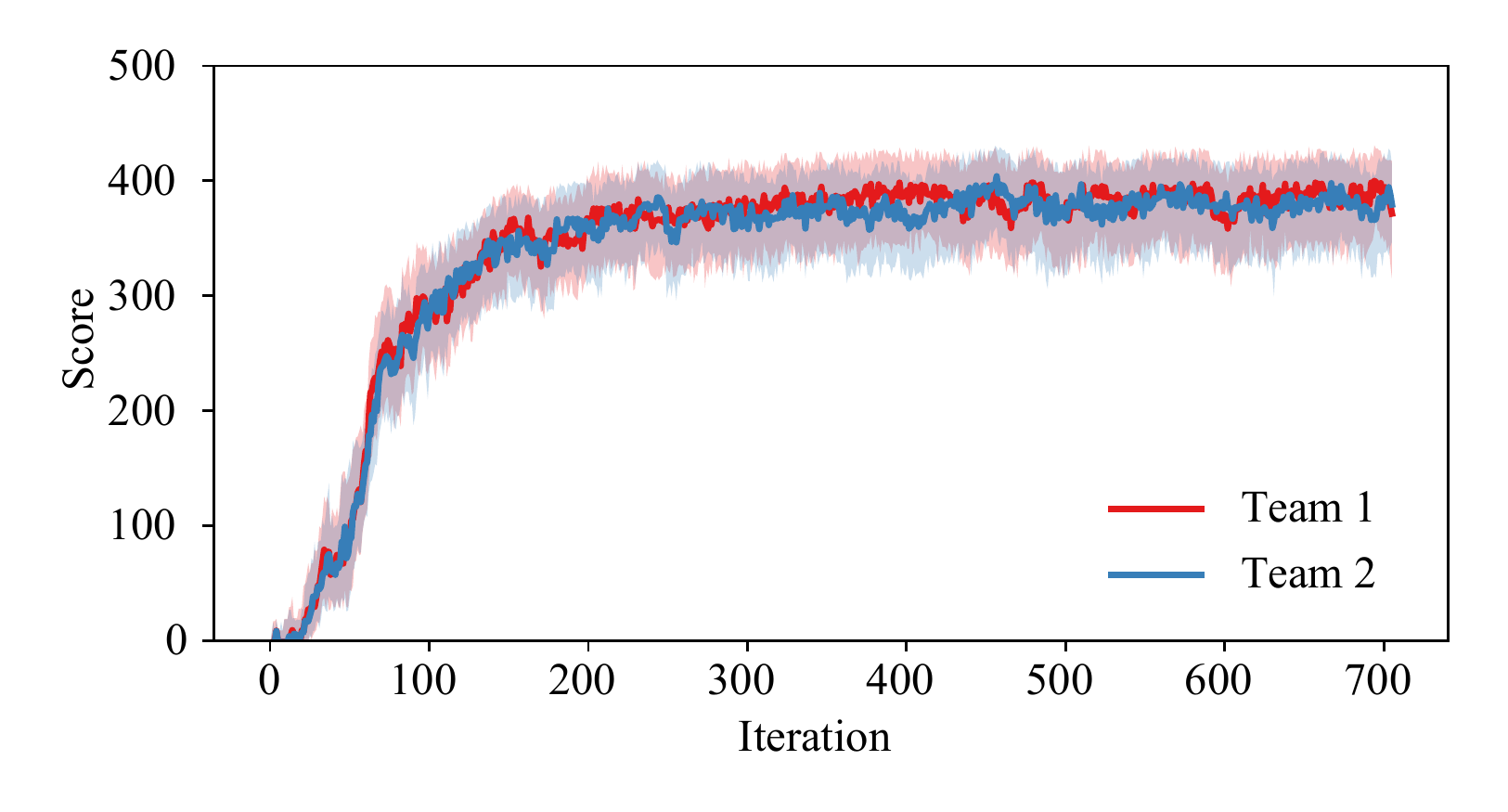}}\\\vspace*{-1em}
  \subfloat[Testing]{\includegraphics[width=1.0\linewidth]{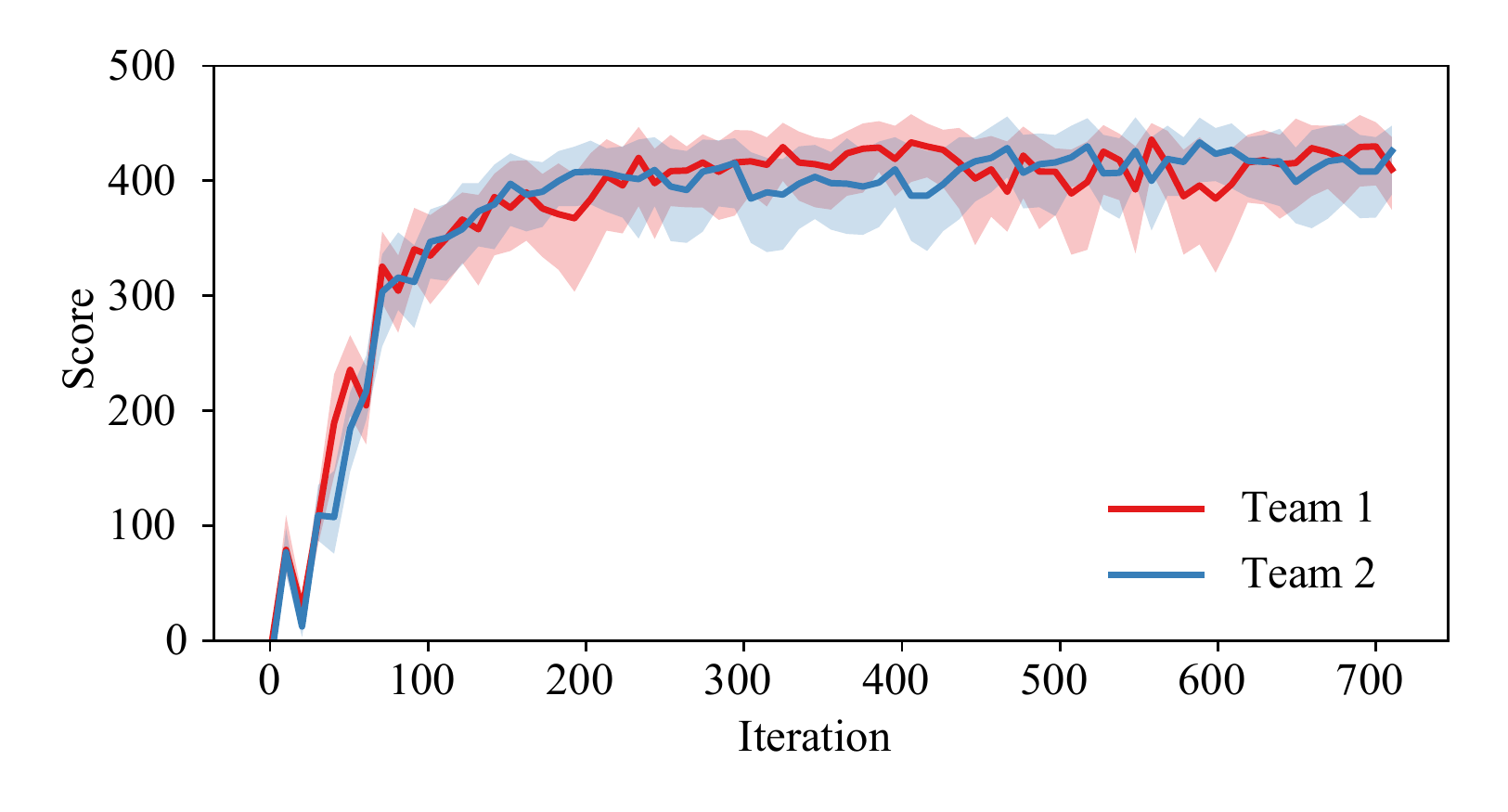}}
  \caption{Scores per iteration for the team competition task.  Top: Scores earned by each team with policies parameterized by randomly peturbed values during training.  Bottom:  Scores earned when testing the trained architectures using updated policy parameters.  Red and blue curves show scores for team 1 and 2 respectively.}
    \label{fig:vs_scores}
\end{figure}
The second scenario we consider is where two teams each equipped with their own unique policies for their agents learn concurrently to defeat their opponent (Figure \ref{fig:versus}).  At each iteration, $N_k=300$ simulations are spawned, each with a different random perturbation, and with each team having a different perturbation.  The updates for each policy are calculated based on the scores received from playing the opponent's perturbed policies.  The result is that each team learns to defeat a wide range of opponent behaviors at each iteration.  We observed that the behavior of the two teams quickly approached a Nash equilibrium where both sides try to defeat the maximum number of opponent aircraft in order to prevent higher-scoring suicide attacks (see supplementary video).  The end result is a stalemate with both teams annihilating each other, ending with tied scores (Figure \ref{fig:vs_scores}).  We hypothesize that more varied behavior could be learned by having each team compete against some past enemy team behaviors or by building a library of policies from which to select from, as frequently discussed by the evolutionary computation community \cite{stanley2004competitive}.
\section{Conclusion}
We have shown that Evolution Strategies are applicable for learning policies with many thousands of parameters for a wide range of complex tasks in both the competitive and cooperative multi-agent setting.  By showing the connection between ES and more well-understood model-based stochastic search methods, we are able to gain insight into future algorithm design.  Future work will include experiments with optimizing mixed parameterizations, e.g. optimizing both neural network weights and PID gains.  In this case, the second-order treatment on non-neural network parameters may be more beneficial, since the behavior of the system may be more sensitive to perturbations of non-neural network parameters.  Another direction of investigation could be optimizing unique policies for each agent in the team.  Yet another direction would be comparing other evolutionary computation strategies for training neural networks, including methods which use a more diverse population \cite{Conti2017}, or more genetic algorithm-type heuristics \cite{Such2017}.

\bibliographystyle{IEEEtran}
\bibliography{iros2018_swarm}
\end{document}